\documentstyle[preprint,aps,eqsecnum]{revtex}
\begin{document}

\preprint{PREPRINT OF IRE 95-11}

\title
{FUNDAMENTAL STEPS OF GROUP VELOCITY \\
FOR SLOW SURFACE POLARITON UNDER    \\
THE QUANTUM HALL EFFECT CONDITIONS}

\author{{\bf IGOR E. ARONOV and NIKOLAI N. BELETSKII}}

\address{Institute for Radiophysics and Electronics, \\
National Academy of Sciences of Ukraine,      \\
12 Acad. Proscura St., Kharkov 310085, Ukraine}

\date{May 31, 1995}

\maketitle

\begin{abstract}
A new type of collective electromagnetic excitations, namely  surface
polaritons (SP) --- in a 2D electronic layer in a high
magnetic field under Quantum Hall Effect (QHE) conditions is predicted.
We have found the spectrum, damping, and polarization of the SP in a wide
range of frequencies $\omega$ and wavevectors $\bf k$.  It is shown that
near the Cyclotron Resonance (CR)
($\omega\sim\Omega=\displaystyle eB/mc$)
the phase velocity of the SP  is drastically slowed down and the
group velocity undergoes fundamental steps defined by the Fine Structure
Constant $\alpha=e^2/\hbar c$.  In the vicinity of a CR subharmonic
($\omega\sim 2 \Omega$)
the negative (anomalous) dispersion of the SP occurs.
The relaxation of electrons in the 2D layer gives rise to a new
dissipative collective threshold-type mode of the SP.
We suggest a method for calculating the kinetic coefficients for
the 2D electronic layer under QHE condition, using the
Wigner distribution function formalism and determine their spatial and
frequency dispersion. Using this method we have calculated the line-shape
of the CR and the d.c. conductance under the QHE condition, which are in good
agreement with experimental data.
\end{abstract}

\draft

\section{{\bf Introduction}} \label{one}
Since the discovery of the Quantum Hall Effect (QHE) \cite{r1,r2,r3}, a
number of authors have investigated the weak damping of collective
electromagnetic waves in 2D electronic layers in a strong magnetic field
$\bf B$ \cite{r3,r4}.
The quantization of the Hall conductivity and the vanishingly small
dissipative (longitudinal) conductivity lead, under the QHE conditions,
to spatial and frequency dispersion in the system and hence
to the generation of an unusually slow collective wave, whose
 dispersion characteristics are also quantized.

 In this paper a new type of collective electromagnetic excitations in
a 2D electronic system under the QHE conditions  is predicted, viz.,
the slow surface polaritons (SP). We calculate the spectrum, damping,
and polarization of that wave in a wide range of frequencies
  $\omega$ and  wavevectors $\bf k$.
The phase velocity of the SP is drastically slowed down near the
principal cyclotron resonance
 ($\omega\sim\Omega$, where $\quad \Omega=eB/mc$  is the cyclotron frequency)
and their group velocity undergoes  fundamental jumps,
whose magnitude is determined by the Fine Structure Constant
 $\alpha=e^2/\hbar c$.
The number of the slow SP modes is conditioned be the magnitude of
the Landau-level filling factor
 $\aleph=\pi\ell^2 n$ (where $\ell=(c\hbar/eB)^{1/2}$  is
the magnetic length, and  $n$ the density of 2D electrons), i.e.,
 by the value of the quantized Hall conductivity.
 In the vicinity of the CR subharmonic
 $(\omega\sim\ 2 \Omega)$,
 the negative (anomalous) dispersion of SP (see Fig.~\ref{Fig.4}) occurs.
 Besides, a new type of SP appears near the CR, which is dissipative
 in nature.  The condition of existence of that additional SP is
 determined by the quantized threshold criterion, which allows
 determining the relaxation frequency at low temperature to an accuracy of
 $\alpha$ (see Eq.~(\ref{Eq34})).

The QHE has been intensely studied with the use of
 methods of modern condensed-matter theory \cite{r3}.
The Integer QHE (IQHE) is thought to be caused by  localization of
electrons in the two-dimensional systems, and the Fractional QHE (FQHE)
is due to electron-electron interaction, which leads to generation of
correlated many-particle ground state \cite{r3} at  distinct fractional values
of
the Landau-level filling factor  $\aleph$.
As this takes place, the paradox lies in the fact that the presence of
"dirt" is the necessary condition for the
  localization-delocalization phase
transition effect; although it is well known that for the observation of
QHE, and, particularly FQHE, the use of  perfect samples with a high
mobility is required.
QHE is one of the problems of the {\it postmodern quantum mechanics} (to use
the term of R.Harris), discussed in four papers in Physics Today in 1993
\cite{r5}.  The technological advances of  the last
decade have succeeded in  fabrication of two-dimensional systems in which
the ballistic (or quasiballistic) transport with large mean free paths can be
realized.

 From the above reasoning we suggest a simple method, which uses almost the
Pauli principle alone, for the description of the kinetics of electrons in
2D systems placed in a strong quantizing magnetic field $\bf B$.
By using the Wigner distribution function \cite{r6} we derive the kinetic
characteristics of a 2D electronic gas under the
QHE conditions and find their
spatial and frequency dispersion.
By means of these results we can adequately describe  the d.c. effects
of IQHE, as well as the line-shape of the CR under the IQHE condition and
hence the dynamics of collective electromagnetic excitations
 under  QHE
conditions.  The article is organized as follows. In Section \ref{two} we
use the Wigner distribution function for  describing the transport
phenomena in a 2DES under QHE conditions. We calculate the conductivity
tensor with a spatial and frequency dispersion. In Section \ref{three} we
present the electrodynamics in 2DES in the high magnetic fields (QHE
effect).  We derive the dispersion relation for  electromagnetic surface
waves and discuss the dispersion, polarization and damping for the
quantized SP in this system. We conclude the paper  with
a brief summary of results and possible applications (Section \ref{four}).

\section{{\bf Transport in the Quantum Hall Effect}} \label{two}

To find the conductivity tensor accounting for the spatial and frequency
dispersion in a 2D electronic gas placed in a high quantizing magnetic
field $\bf B$ (under QHE conditions) oriented normally to the 2D layer (see
Fig.~\ref{Fig.1}), we will apply the Wigner distribution function
\cite{r6,r7,r8}:  \begin{equation} f_{{\bf p}}^W({\bf r})=\int d{\bf
r}^{^{\prime}} Tr\{\hat{\rho }\, exp[-i({\bf p}+{e\over c}{\bf A} ({\bf
r})){\bf r}^{^{\prime}}] \psi^{+}({\bf r}- {\bf r}^{^{\prime}}/2)\psi
({\bf r}+{\bf r}^{^{\prime}}/2)\}.
\label{Eq1}
 \end{equation}

Here $\hat{\rho}$  is the statistical operator of the system;
$\psi^{+}({\bf r})$ and $\psi ({\bf r})$ are the Fermi operators of
generation and annihilation respectively, of particles  at  point  $\bf r$;
$\bf A$ the vector-potential of the electromagnetic
field. In the case when the scale size of the spatial inhomogenity
 exceeds both the radius of interaction between the particles and the de
Broglie electron  wavelength, the kinetic equation for the Wigner
 distribution function Eq.~(\ref{Eq1}) takes the form \cite{r6,r7,r8}
equivalent to the classical kinetic equation:
\begin{equation} {{\partial
f_{{\bf p}}^W}\over {\partial t}}+{\bf v}{{\partial f_{{\bf p}}^W}\over
{\partial {\bf r}}}+e\{{\bf E}+{1\over c}[{\bf v} ,{\bf B}]\}{{\partial
f_{{\bf p}}^W}\over {\partial {\bf p}}}=\hat { I}\{f_{{\bf p}}^W\}.
\label{Eq2}
\end{equation}
Here $\bf E$  and $\bf B$ are the electric field and the magnetic induction
vectors;
$e$ the electron charge, and $\bf v$ the velocity of conduction electrons.

In the case under consideration, if the 2D electronic system is infinite
in the $xy-$plane (see Fig.~\ref{Fig.1}), then the typical scale of
inhomogenity is the wavelength $k^{-1}$ of the collective electromagnetic
wave. Thus, the existence criteria for Eq.~\ref{Eq2} are $k \ll n^{1/2}$, (
since with weak screening  $n^{-1/2}$ is the characteristic length
interaction between the particles) and  $k\ell \ll 1$ (since in a strong
magnetic field the magnetic length $\ell=\displaystyle({c\hbar \over
eB})^{1/2}$ represents the de Broglie wavelength of electrons). The collision
integral, $\hat {I}\{f_{{\bf p}}^W\}$, differs essentially from the classical
collision integral, since the quantum transitions accounted for by, $\hat
{I}\{f_{{\bf p}}^W\}$, reflect the character of statistics, obeyed by the
particles, and the distinction of the Wigner distribution function from
the classical one \cite{r7}.  The equilibrium Wigner distribution
function sets the collision integral, $\hat {I}\{f_{{\bf p}}^W\}$, to
  zero. The equilibrium Wigner function can be expressed via its value for
an equilibrium ensemble of quantum states of an electron in a magnetic
field $\bf B$. By using the definition Eq.~(\ref{Eq1}) and substituting
the wavefunctions of an electron in an electromagnetic field into
Eq~(\ref{Eq1}), we obtain for the spinless electrons \cite{r7,r8}
\begin{equation}
 f_0(\epsilon)=\sum_{s=o}^{\infty}n_F\left [{{\hbar\Omega
(s+{ 1\over 2})-\mu}\over T}\right ]\Gamma_s({{\epsilon}\over
{\hbar\Omega}}).
\label{Eq3}
\end{equation}
$$ \Gamma_s(x)=2(-1)^s
exp(-2x)L_s^{(0)}(4x), $$ $$ n_F(x)=(1+e^x)^{-1}.  $$ Here $\epsilon
=\displaystyle{{p^2}\over {2m}}$ is the energy of 2D electrons and
$L^{(0)}_s(x)$ the Laguerre polynomial.  If we replace the summation over
$s$ by integration, then for $\hbar\Omega\ll T$ ($T$ is the temperature)
Eq.~(\ref{Eq3}) transforms into an equilibrium Fermi distribution function
($\mu$ is the chemical potential).

With the knowledge of the equilibrium Wigner distribution function we can
describe all the thermodynamical relations.  In this paper we will consider
the 2DES when the chemical potential $\mu$ is constant over the entire
system.  The relation between the
 electron density $n$ and the chemical
potential $\mu$ in a strong magnetic field can be found, as usual,
from the normalization condition.  The density $n$ of 2D electrons is the
average of $f_0(\epsilon)$ Eq.~(\ref{Eq3}).  As a result  we obtain
after the averaging
\begin{equation} \aleph
=\pi\ell^2n=\sum_{s=0}^{\infty}n_F\left [{{\hbar\Omega (s+{1\over
2})-\mu}\over T}\right ].
\label{Eq4}
 \end{equation}

It is shown that the Landau-level filling factor $\aleph$
assumes only positive integer values if $T\ll\hbar\Omega ,\mu$.
  The fact that the filling factor $\aleph$ can assume
 only  integer values leads to IQHE. However, Eq.~(\ref{Eq4}) contains a
 contradiction. Indeed, why should the ratio of  two independent values
 which the electron density $n$ and the magnetic induction $\bf B$ in the
 sample are take only  integer values? It is well-known that the
 mean value of a microscopic magnetic field is the magnetic induction $\bf
 B$.  The value of $\bf B$ in the sample should be found from the formula
 ${\bf H}={\bf B}-4\pi {\bf M}({\bf B})$, where $\bf H$ is the external
 magnetic field and ${\bf M}({\bf B})$ the magnetic moment. If the
 temperature is not too low, then $M({\bf B})\ll B$ and the magnetic
 induction value  $B$  differs but slightly from $H$.  As
temperature is decreased, the amplitude of oscillations of
the magnetic moment ${\bf M}({\bf B})$ increases and a situation appears,
when the regions of H-values corresponding to three various values of $\bf
B$, show up.  This ambiguity indicates an instability of states similar to
that taking place on the Van-der-Waals curve of the equation of state
\cite{r9,r10,r11}.
In other words, at such values of the external magnetic fields
diamagnetic phase transitions take place in the system, at which an
inhomogeneous state ( domain type and(or) periodic structures) appear in
the system, when the magnetic induction  $B$ and the electron density $n$
become coordinate-dependent functions.  In the vicinity of  such phase
transitions and at the inhomogeneous states, the scaling-type dependencies of
the conductivity on the magnetic field and  singularities at the fractional
values of the Landau-level filling factor $\aleph$ should appear due to the
scaling-type invariance (see \cite{r9,r10,r11}).  This kind of
 behavior should result in singularities in the Hall conductivity and in a
 longitudinal (dissipative) conductivity of the 2D electronic gas at the
 fractional values of the filling factor $\aleph$, i.e.,  to FQHE.  In
 particular, when the magnetic induction and the electron density are
 characterized by an
 inhomogeneous structure , the splitting of states and the additional gaps
 might  originate in the electron spectrum, and  the spectrum degeneracy in
the orbit centre  coordinate  in the magnetic
field may become removed.

 Besides,  an additional drift of
electrons \cite{r12} can arise in a weakly inhomogeneous magnetic
 field which   can give an additional contribution to the Hall current.
 However, far apart from the diamagnetic phase transition
 \cite{r9,r10,r11}  equilibrium states exist; the electron density is
independent of coordinates, and the magnetic induction $\bf B$ assumes a
 sufficient value to satisfy relation Eq.~(\ref{Eq4}), and the Landau-level
filling factor is an  integer value, i.e., the IQHE condition is met.
 Thus, in this paper we will analyze the electron kinetics under the IQHE
 condition.

 The form of the electron-phonon and electron-impurity collisions
 integral Eq.~(\ref{Eq3}) is too complicated \cite{r7}. However, we will
 consider here the effects determined by the linear response to the electric
 field. In this case the distribution function, $f_{\bf p}^W$, can be found
 in an  approximation linear in the external field, $\bf E$.  It is
 well-known \cite{r7} that for  sample with a high electron mobility
 the collision integral can be represented in the $\tau$--approximation,
 where the mean free path time $\tau$ is determined by the momentum
 relaxation frequency, being a function of the electron energy,
 $\epsilon$.  In other words, for this quasiballistic regime we can find
 the Wigner distribution function in the form:
  \begin{equation} f_{{\bf
  p}}^W={\bf f}_0(\epsilon )+f_1.
 \label{Eq5}
  \end{equation}
  where $f_0(\epsilon )$
is the equilibrium distribution function in a high magnetic field
 Eq.~(\ref{Eq3}), and $f_1(t,{\bf p},{\bf r})$ the correction to the
Wigner distribution function, which is determined by the electric field
$\bf E$.  The collision integral, $\hat{I}\{f_{{\bf p}}^W\}$, will be
written as
\begin{equation} \hat {I}\{f_{{\bf p}}^W\}=-\nu(\epsilon )f_1.
\label{Eq6}
\end{equation}

In the general case we will assume the electric field to be
a  function of coordinates and time.
Then for the Fourier-transform of the current density, $\bf j$,
and the electric field, $\bf E$, one obtains
the linear relations
\begin{equation}
j_{\alpha}(\omega ,{\bf k})=\sigma_{\alpha\beta}(\omega ,{\bf k} )\,{\cal
E}_{\beta}(\omega ,{\bf k});
\label{Eq7}
\end{equation}
\begin{equation} E_{\alpha}({\bf
  r},t)= \int{{d^2\!k\,d\omega}\over {(2\pi )^3}}{\cal E}_{
  \alpha}({\omega,\bf k})\, e^{i({\bf k}{\bf r}-\omega t)};
\label{Eq8}
\end{equation}
\begin{equation} {\cal
E}_{\alpha}(\omega ,{\bf k})=\int d^2\!k\,d\omega E_{\alpha}(\omega,{\bf
k})\,e^{i({\bf k}{\bf r}-\omega t)}.
\label{Eq9}
\end{equation}

The Fourier-transform of the conductivity tensor,
$\sigma_{\alpha\beta}(\omega ,\bf k)$, accounting for the
spatial and frequency dispersion can be found by using the Wigner distribution
function Eqs.(\ref{Eq2}),(\ref{Eq3}),(\ref{Eq5}). Thus, in the general
case we have
 \begin{equation} \sigma_{\alpha\beta}(\omega ,{\bf
k})={{2e^2}\over h}^{}\sum^{\infty}_{ s=0}n_F\left\{{{\hbar\Omega
 (s+{1\over 2})-\mu}\over T}\right\} \int_0^{\infty}d\xi{{(\pi^2\xi
/2)}\over {sh[\pi\gamma (\xi )]}} {{\partial\Gamma_s(\xi )}\over
{\partial\xi}}D_{\alpha\beta}(\xi )
\label{Eq10}
 \end{equation}

Here the tensor components,  $D_{\alpha\beta}(\xi )$, are given by
\begin{eqnarray}
 D_{xx}(\xi )&=& m_1(\xi )+m_2(\xi )+2m_3(\xi
)\cos 2\beta ;\nonumber \\
 D_{yy}(\xi )&=& m_1(\xi )+m_2(\xi )-2m_3(\xi )\cos
2\beta ;\\  \label{Eq11}
D_{xy}(\xi )&=&-im_1(\xi )+im_2(\xi )+2m_3(\xi )\sin 2\beta
;\nonumber \\ D_{yx}(\xi )&=&im_1(\xi )-im_2(\xi )+2m_3(\xi )\sin 2\beta.
\nonumber
 \end{eqnarray}

The functions  $m_1(\xi )$, $m_2(\xi )$ and $m_3(\xi )$
are defined by the expressions:
\begin{eqnarray}
 m_1(\xi )&=&J_{-i\gamma -1}(z)J_{i\gamma +1}(z);\nonumber \\
 m_2(\xi )&=&J_{-i\gamma +1}(z)J_{i\gamma -1}(z); \\  \label{Eq12}
 m_3(\xi )&=&J_{i\gamma +1}(z)J_{-i\gamma +1}(z), \nonumber
\end{eqnarray}
where $z=\sqrt {2\xi}kl$; $\gamma =\displaystyle{{\nu (\xi\hbar\Omega
)-i\omega}\over { \Omega}}$; $k=\sqrt{k^2_x+k^2_y}$; $\cos
\beta=\displaystyle{{k_x}\over k}$; and $\sin \beta=-\displaystyle{{k_y}\over
k}$.

Note that the Fourier-transform of conductivity,
Eqs.(\ref{Eq10})-(\ref{Eq12}) contains terms proportional to  $\cos 2\beta$
and $\sin 2\beta$, whose appearance results from the two-dimensionality of
the electronic gas. They violate the symmetry of the kinetic coefficients
for the Fourier-transform of, $\sigma_{\alpha\beta}(\omega ,{\bf k})$.
 Such polarizable terms may be important for  the polarization of
 electromagnetic eigenoscillations in 2D electronic systems. The
 integration over ${\bf d}^2{\bf r}$ re-establishes the symmetry.

One can easily see that  generally  the conductivity tensor
experiences resonance oscillations, viz., of the cyclotron-resonance
type in  the case of a strong spatial dispersion (when $k\ell^*\gg 1$,
where $\ell^*$ is the mean free path) on the multiple harmonics
(subharmonics of the CR when $\omega =s\Omega$, $s=1,2,\ldots$).
However, the case of a strong spatial dispersion can be practically realized
 only at  very high frequencies. As is easy to see from Eq.~(\ref{Eq3})
 and from the expression for the conductivity tensor the strong
 quantization in a magnetic field "destroys the Fermi surface" and the
conductivity is due to  all electrons with various energies.  Thus, the
frequency dispersion of the conductivity (and the spatial dispersion as
  well) is in essence defined by the form of the function $\nu = \nu
(\epsilon)$. At low frequencies, when $\omega\leq\nu(\epsilon)$, it might
be an "indicator" of the function $\nu(\epsilon)$. It is also clear that
the lineshape of the CR can be essentially dependent on the function
$\nu=\nu(\epsilon)$ \cite{r10,r13}.  The form of the longitudinal d.c.
  conductivity ($\sigma_{xx}=\sigma_{yy}$, when $\omega=0$, $k=0$) is also
drastically dependent on the type of the function $\nu=\nu(\epsilon)$
\cite{r13}.  The lineshape of the CR can be changed if the electron
effective mass $m=m(\epsilon)$ depends on the electron energy $\epsilon$,
when the dispersion law of conduction electrons differs from the quadratic
 one \cite{r14}.  In this paper we will assume that the mobility is very
 high and we will find the conductivity when the relaxation frequency is
 an effective constant, i.e., $\nu$=const \cite{r15}. We will
 consider the most realistic case, when the spatial dispersion is
 sufficiently weak, i.e.,
 \begin{equation} k\ell\ll 1.
 \label{Eq13}
  \end{equation}

Then the Fourier-transform of the conductivity tensor can be obtained
in the form:
\begin{equation}
 \sigma_{\alpha\beta}={2e^2\over h}{\aleph\over
{1+\gamma^2}}\left\{ B_{\alpha\beta}-{(k\ell )^2\over {2\gamma}}
\left(1+{\aleph\over 2}\right)
C_{\alpha\beta}\right\},
 \label{Eq14}
\end{equation}

where

\begin{eqnarray}
 B_{xx}&=&B_{yy}=\gamma ; \qquad \qquad B_{xy}=-B_{yx}=1; \nonumber \\
 C_{xx}&=&a+\cos 2\beta ;\qquad C_{yy}=a-cos2\beta ; \\ \label{Eq15}
 C_{xy}&=&-b-sin2\beta ; \quad   C_{yx}=b-\sin 2\beta .\nonumber
\end{eqnarray}
$$ a={{2(\gamma^2+2)}\over {\gamma^2+4}}; \qquad b={{6\gamma}\over
{\gamma^2+4}}.  $$

Let us summarize the formulas for the resistance tensor in the d.c. case,
when $\omega=0$, $k=0$ (IQHE)
\begin{equation}
\rho_{xx}={{\sigma_{xx}}\over {\sigma^2_{xx}+\sigma^2_{xy}}}={h\over {
2e^2}}{{\gamma}\over {\aleph}};
 \label{Eq16}
\end{equation}
\begin{equation}
\rho_{xy}={{\sigma_{xy}}\over {\sigma^2_{xx}+\sigma^2_{xy}}}={h\over {
2e^2}}{1\over {\aleph}}.
 \label{Eq17}
\end{equation}

Fig.~\ref{Fig.2}  shows the graphs of  $\rho_{xx}$ and $\rho_{xy}$ as
functions of the magnetic field ($\omega=0$)
calculated by Eqs.~(\ref{Eq16}) and (\ref{Eq17}).
(The chemical
potential $\mu=20 meV$, the mean frequency of a momentum relaxation
$\nu=10^{12}s^{-1}$, for the temperature $T=100 mK$).
 As can be seen,
 Eqs.(\ref{Eq16}) and (\ref{Eq17}) are good enough to describe the
classical picture of IQHE \cite{r3,r16}, even at $\nu$=const .
Eq.~(\ref{Eq14}) and Eq.~(\ref{Eq15}) show that if $\nu$=const then the
relation $\rho_{xx}$/$\rho_{xy}=\gamma=\nu/\Omega$ should exist, which
can be
observed under the IQHE condition. The deviation from this
simple relation demonstrates the energy dependence of the relaxation
frequency, $\nu=\nu(\epsilon)$ \cite{r16}.  Here we will show the
lineshape of the CR (the high-frequency absorption $\sim {\rm
Re}\,\sigma_{xx}$) for various frequencies , $\omega$, as a function of
the magnetic field (see Fig.~\ref{Fig.3}).  Obviously, the lineshape of
the CR is highly sensitive to the CR position.  In the case when the line
centre  is located at the centre of the IQHE plateau, it shows kinks at
the points, where the jumps of QHE occur.  Fig.\ref{Fig.3}(a) shows the
lineshape of cyclotron resonance in 2DES calculated for GaAs/GaAlAs  by
formula Eq.~\ref{Eq14}, for the temperature $T=50 mK, \quad (k \rightarrow
0)$ and $\nu=2\cdot 10^{11} s^{-1}$.   The wide resonance line, which
captures the several steps of QHE for the rather low frequency $\omega$,
when the CR takes place at the quite small value of magnetic field. It is
shown that the additional structure of the line, which came about from the
interference of singularities of the CR and QHE.  If the centre of CR line
 is located near the QHE step, then the amplitude of the CR is increased.
Fig.~\ref{Fig.3}(b) shows
the narrow resonance lines, which take place for the higher
frequencies $\omega$ located in the range of one step of the QHE.
It is shown that the structure of line, whose centre is in the centre of
the Hall plateau, exists on the wings of line at the values of
magnetic field $\bf B$  corresponding to the jump of the Landau-level
filling factor $\aleph$. The amplitude of CR line is increased, when the
center of CR line is located at the point of jump of the Landau-level
filling factor $\aleph$ (The lines for the frequencies $\omega_3=1.00\cdot
10^{13} s^{-1}, \omega_5=1.35\cdot 10^{13} s^{-1}$).  Such features of the
CR line were observed by a number of authors \cite{r17}.

\section{{\bf Electrodynamics of 2DES under the QHE condition}}
\label{three}
The propagation of electromagnetic waves  trough systems with a 2D
electronic gas in the dielectric environment, placed in a strong magnetic
field (see Fig.~\ref{Fig.1}), is described by the Maxwell equations for the
scalar and vector potentials in the Lorentz gauge, viz.
 \begin{equation}
\nabla {\bf A}+{{\epsilon}\over c}{{\partial\varphi}\over {\partial t}}=0.
\label{Eq18}
\end{equation}
The potentials satisfy the usual wave equation  \cite{r18,r19}
\begin{equation}
\left[\nabla^2-{{\epsilon}\over {c^2}}\left({{\partial}\over {\partial
t}}^{}\right)^2\right]\varphi ({\bf r},t)=-{{4\pi}\over {\epsilon}}
\rho_{tot}({\bf r},t),
\label{Eq19}
\end{equation}
\begin{equation}
\left[\nabla^2-{{\epsilon}\over {c^2}}\left({{\partial}\over {\partial
t}}^{}\right)^2\right]{\bf A}({\bf r},t)=-{{4\pi}\over c}{\bf j}_{
tot}({\bf r},t).
\label{Eq20}
\end{equation}
These values are related to the fields in Eq.~(\ref{Eq2}) as
\begin{equation}
{\bf E}=-\nabla\varphi -c^{-1}{{\partial {\bf A}}\over {\partial
t}};\quad {\bf B}={\rm rot}{\bf A.}
\label{Eq21}
\end{equation}

Here
$\rho_{tot}=\rho_{ex}+\rho$ is the total charge density in the system
and  ${\bf j}_{tot}={\bf j}_{ex}+{\bf j}$ the total
current density; $\rho_{ex}$ and ${\bf j}_{ex}$ are the external charge
and current densities, respectively. In the system under consideration
$\rho\sim\ {\bf j} \sim\delta (z)$ \cite{r19} , so that the charges
and currents exist only in the 2D electronic layer, and the external currents
and charges in the system are absent: $\rho_{ex}={\bf j}_{ex}=0$.
In this case the potentials ${\bf A}$ and $\varphi$ can be
found from the homogeneous  equation set Eqs.(\ref{Eq19}), (\ref{Eq20}).
Using the Fourier transformation Eq.~(\ref{Eq7}) and taking into account
 ${\bf j}(\omega,{\bf k},z)= {\bf j}(\omega,{\bf k},0)\delta(z)$,
we obtain:
\begin{equation}
{\bf A}(\omega,{\bf k},z)={\bf A}_0(\omega,{\bf
k})e^{-p|z|},
\label{Eq22}
\end{equation}
where  $p=\displaystyle\sqrt{k^2-{\omega^2\over c^2}\varepsilon},\quad {\rm
Re}\,p>0$.  In other words, the system supports eigenoscillations in the form
of a surface wave pressed up against the 2D electronic layer, (see
Fig.~\ref{Fig.1}).  In this case the component $A_z=0$, while the scalar
   potential can be found from the Lorentz gauge (\ref{Eq17}):  ${\bf
   kA}=\displaystyle{\omega\varepsilon\over c}\varphi$.  The current density
  in the 2D electronic layer can be represented in the form
\begin{equation}
 {\bf j}_{\alpha}(\omega,{\bf k})=i{{\omega}\over
c}\sigma_{\alpha \beta}(\omega,{\bf k})\left[A_{\beta}(\omega,{\bf
k},0)-{{c^2}\over {
\varepsilon\omega^2}}k_{\beta}k_{\gamma}A_{\gamma}(\omega,{\bf
 k},0)\right].
\label{Eq23}
 \end{equation}
Thus, the dispersion relation is found from the condition
\begin{equation}
{\bf A}(\omega,{\bf k},0)={{2\pi}\over {cp}}\,{\bf j}(\omega,{\bf k},0).
\label{Eq24}
\end{equation}
This dispersion relation takes the form:
\begin{equation}
{\bf D}=Det \left\{\delta_{\alpha\beta}-{{2\pi i\omega}\over {c^2p}}\left
[\sigma_{\alpha\beta}(\omega,{\bf k})-{{c^2}\over {\epsilon\omega^2}}
\sigma_{\alpha\gamma}(\omega,{\bf k})k_{\gamma}k_{\beta}\right]\right
\}=0,
\label{Eq25}
\end{equation}
where $\delta_{\alpha b}$ is the Kronecker delta.

Fig.~\ref{Fig.4}  presents the dispersion curves of the SP propagating in
the system of Fig.~\ref{Fig.1}.
The dispersion curves for the surface polariton on the boundary
of 2DES calculated for the various values of the Landau-level filling
factor $\aleph$ ($\aleph=1; \aleph=5$, and $\aleph=10$). 2DES is realized
by heterostructure GaAs/GaAlAs, the effective mass $m=0.068m_0$,
$\varepsilon=12$.
The $y$--axis gives the real part of frequency, and the $x$--axis gives
the wave number.
 It is seen that the spectrum of the SPs  is
gapless at  low frequencies ($\omega \gg \Omega$) and they exist
both in the low-frequency region $\omega<\Omega$ and in the
high-frequency region $\omega>\Omega$.
In the low-frequency region, far away from the CR, the phase
velocity of the SP is close to the light velocity
 $v_d=c/\sqrt{\varepsilon}$ in the dielectric, which surrounds the 2D
electronic layer.  In the high-frequency region and in  the vicinity of the
principal CP ($\omega\sim\Omega$) the phase velocity of the SPs
drastically decreases and they are transformed into  slow waves. In the
frequency range $\Omega<\omega<2 \Omega$, where $\ell^{-1}\gg k
\gg \displaystyle {\omega \over c} \sqrt {\varepsilon}$, one can neglect the
retardation effect and the spatial dispersion in the conductivity tensor
of Eq.~(\ref{Eq14}).
The dispersion relation can be brought to the form
   \begin{equation}
  \omega^2=\Omega^2+{2\pi v_nk\Omega \over \varepsilon},
\label{Eq26}
\end{equation}
   where
$v_n=\displaystyle{2e^2\over h}\aleph$.  It is easy to see  that with
$\Omega \gg {\displaystyle v_nk \over \varepsilon}$ ($\aleph \sim 1$) the
dispersion law of the SP is linear ($\omega\sim k$), while in the opposite
case $\Omega \gg \displaystyle {v_nk \over \varepsilon}$ (or when $\aleph \gg
1$) the dispersion law is of a square-root type  ($\omega\sim\sqrt k$).

Near the CR ($\omega\sim\Omega$) the retardation effect cannot be
neglected and the dispersion law of the SP becomes
\begin{equation}
\omega=\Omega +\Omega{{v_n}\over
c}\left \{{{\pi\Omega}\over {cp_1}}\left ({{p^2_1c^2}\over
{\epsilon\Omega^2}}-1\right)+{{2\pi^2v_n}\over {\epsilon c}}\right\}-i\nu .
\label{Eq27}
\end{equation}

 Here $p_1=\displaystyle\sqrt {k^2-{\Omega^2\over {c^2}}\varepsilon}$.
 The  value of the relative deceleration of the SP is determined by the
  Fine Structure Constant $\alpha$.
The group velocity, ${\bf v}_g=
\displaystyle{{\partial\omega}\over {\partial {\bf k}}}$,
of the SP undergoes the fundamental steps in  the vicinity of the CR :
\begin{equation}
{{v_g}\over {v_d}}={{2\sqrt 2\alpha\aleph}\over {\sqrt
{\varepsilon}}}.
\label{Eq28}
\end{equation}
In other words, the deceleration of the wave
near the CR is considerable, and the reason for the quantization of the
group velocity is the quantization of the Hall conductivity,
i.e., in fact the system possesses a fundamental parameter of the
velocity dimension,  the conductance quantum
$\displaystyle 2e^2\over h$. At the point of the CR
the character of conductivity is changed and it becomes  imaginary,
i.e. reactive one, and therefore the
conductance becomes nondissipative, and the slow wave (the slow SP)
appears. With a further increase of  frequency $\omega$,
near the doubled CR ($\omega\sim 2\Omega$), the spatial dispersion effects of
 conductivity Eq.~(\ref{Eq14}) becomes noticeable, the group velocity
changes its sign and takes negative values. In this region the SP shows
the anomalous (negative) dispersion.

The SP spectrum  near the CR subharmonic  ($\omega\sim
2\Omega$) is obtained in the form
\begin{equation}
 \omega =2 \Omega -\Omega{b\over
d}\left({{k\ell}\over 2}\right )^2\left(1+{{\aleph}\over 2}\right)-i\nu.
\label{Eq29}
\end{equation}
Here
 \begin{equation} b={{v_n}\over c}\left({{4\pi}\over
 3}\right)\left({{\Omega}\over { cp}}\right)\left(1-{{p_2^2c^2}\over
{4\varepsilon\Omega}}\right);
\label{Eq30}
\end{equation}

$ d=1+2b$ and $ p_2=\displaystyle\sqrt {k^2-{{4\Omega^2}\over
{c^2}}\varepsilon}$.
At $\omega>2\Omega$ the SP propagates
through the system at a velocity  close to that of
the SP far away from the doubled CR (see Fig.~\ref{Fig.4}).
It can be see from Eq.~(29) that the spectrum of the SP is "strongly
pressed" to the line of the CR subharmonic  ($\omega=2\Omega$).
The relative dispersive width of the SP has a scale of the small parameter
$\displaystyle \hbar\Omega/mc^2\ll 1$ near the CR. This dispersion curve
(see the cut-in in the Fig.~\ref{Fig.4}) of the SP  starts near the
fundamental mode of the light line ($\omega=kv_d$), then it is branched
   (the number of branches is equal to the Landau-level filling factor
   $\aleph$) and the group velocity is quantized in the same way as near
the principal CR ($\omega=\Omega$). But the group velocity is very low
$v_g\sim v_d(\hbar\Omega/mc^2$), in the wavenumber region where the
dispersion curve is "strongly pressed" to the line $\omega=2\Omega$. The
spectrum of the SP "slides" near the line $k\simeq \varepsilon\Omega/v_n$.
At such values of $k$ the spectral curve is  detached from the line
$\omega=2\Omega$ (see Fig.~\ref{Fig.4}).  The relative attenuating rate of
the SP is of the order $\nu/\omega$ and is small for samples with a high
 electron mobility.  At low frequencies ($\omega\ll\Omega$) the SP is a
 wave of the TE type, changing in to TM at $\omega>\Omega$.  The SP
polarization  is defined by the following expression:
 \begin{equation}
  {\bf A}= A_0\left(\begin{array}{c}1\\ q \\ 0\end{array}\right)\,
e^{-p|z|}\,e^{i({\bf kr}- \omega t)},
\label{Eq31}
 \end{equation}

The vector potential components  are interrelated as
$A_y=qA_x$, $A_z=0$, the scalar potential $\varphi$ is given by
the Lorentz gauge (\ref{Eq18}). The polarization parameter $q$ is
\begin{equation}
q=\{1-{{2\pi i\omega}\over {c^2p}}[\sigma_{xx}(1-{{c^2k_x^2}\over {
\varepsilon\omega^2}})-{{c^2}\over {\omega^2}}\sigma_{xy}k_xk_y]\}
/\{\sigma_{xy}(1-{{c^2k_{^{}y}^2}\over {\varepsilon\omega^2}})-{{c^
2}\over {\varepsilon\omega^2}}\sigma_{xx}k_xk_y\}.
\label{Eq32}
\end{equation}
It is easy to see that the SP polarization  under the QHE condition
is sensitive to the terms $\sin(2\beta)$ and $\cos(2\beta)$ of
Eqs.(\ref{Eq14}, and \ref{Eq15}). This leads to  rotation of
polarization plane as a function of the magnetic field {\bf B}.
Far  from the CR, where the  spatial dispersion  of conductivity
is negligible, the polarization parameter $q$ takes the simpler form
\begin{equation}
q={cp \over \omega}\{-i {c \over v_n}{(1+\gamma^2) \over 2 \pi}+
{cp \over \varepsilon \omega} \gamma\}.
\label{Eq33}
\end{equation}
It is evident that the polarization parameter $q$ is quantized
due to the Hall quantization.

In the system under consideration (see Fig.~\ref{Fig.1})  another SP
 exists, which appears near the CR ($\omega \sim \Omega$). This SP is a
dissipative-type wave.
The surface wave exists when ${\rm Re}\,p>0$.
When the relaxation frequency $\nu\neq 0$, the frequency $\omega$
is a complex value. A straightforward analysis shows that the
additional (dissipative) SP mode is practically nondispersional and the
condition for its existence is determined by the threshold
condition
 \begin{equation} {{\nu}\over {\Omega}}>2 \alpha{{\aleph}\over
   {\sqrt {\varepsilon}}}.
\label{Eq34}
 \end{equation}
 The dispersion curve of the
additional (dissipative) SP is shown in Fig.~\ref{Fig.5}.

Thus, the observation of such a wave specifies the relaxation frequency to an
 accuracy up to the Fine Structure Constant $\alpha$.

Fig.~\ref{Fig.5}(a) ($\aleph=1;\quad \nu/\Omega=0.01$)
shows that at such values of the relaxation frequency
$\nu$ the Additional Surface Polariton (ASP) is appears, which has
an endpoint of the spectrum defined by the condition ${\rm Re}\,p=0$.
The SP damping  increases  sharply near the CR, where the SP drastically
decelerates. The damping of the ASP is sharply decreased and approaches
zero, when it tends the endpoint of the spectrum. The ASP exists
on the left side of  the "light line" $\omega=kv_d$, being   in fact the
delocalization wave because it is "weakly pressed" to the 2DES. The SP is
the proper surface wave, which is "strongly pressed" to the 2DES  far from
the principal mode (the "light line") $\omega=kv_d$.

Fig.~\ref{Fig.5}(b) ($\aleph=1;\quad \nu/\Omega=0.2$)
shows the spectra of the ASP and SP, when $\nu$ is
 increased.  At such values of $\nu$ the spectrum is essentially modified.
A gap opens in the spectrum. This gap is brought about by the endpoint in
 the spectrum of the low frequency  SP ($\omega<\Omega$) and the blending
of the ASP with the decelerated SP. The low frequency SP becomes a
completely delocalized wave, it is practically not connected to the 2DES.
The damping of the new blended mode (ASP and SP) has a fixed value
$\nu/\Omega$ in the over a wide range of the wave numbers $k$.

Fig.~\ref{Fig.5}(c) ($\aleph=5;\quad \nu/\Omega=0.1$)
 shows the spectral curves for lower  magnetic fields
(when the Landau-level filling factor is $\aleph=5$). The ASP is blended
with the decelerated part of the SP near the principal CR. The low
frequency part of the principal mode of the SP ($\omega=kv_d$) is
separated from the ASP, and its spectrum (${\rm Re}\, \omega
=\omega^{'}=\omega(k)$) ends at the point  where ${\rm Re}\, p=0$. The
 principal mode $\omega=kv_d$ is "weakly pressed" to the 2DES for  all the
values of $\Omega$ and $k$ because ${\rm Re}\,p \ll {\rm Im}\,p$. But the
 decelerated part of the SP and ASP (the upper curve in \ref{Fig.5}(c))
are "strongly pressed" to the 2DES, since for this part of the spectrum
 ${\rm Re}\,p \gg {\rm Im}\,p$. In other words, the slow SP and ASP are
proper surface waves.

Fig.~\ref{Fig.5}(d) ($\aleph=5;\quad \nu/\Omega=0.2$)
shows the picture of spectral curves, when the
relaxation frequency $\nu$ is greater than that of \ref{Fig.5}c. The
endpoint for the principal SP mode  ($\omega=kv_d$) is moved down off the
ASP and slow SP spectral curve. In that picture only the real parts of the
spectral curves cross,  while the imaginary parts $\omega^"$ of
the frequencies assume different  values \cite{r20}.

The spectral picture of the ASP  changes crucially
at  larger values of the Landau-level filling factor (see
Fig.~\ref{Fig.5}(e), $\aleph=10 \quad \nu/\Omega=0.1$). First, the
curves of the ASP and slow SP are separated, second the ASP curve acquires
an endpoint of the spectrum (${\rm Re}\,p=0$).  It is significant that the
spectrum shows an anomalous (negative) dispersion near the CR and an
endpoint of ASP. At high values of the relaxation frequency (see Fig.~
\ref{Fig.5}(f),
$\aleph =10; \nu /\Omega =0.2$) the picture of the dispersion curves is of
the same kind as in Fig.~\ref{Fig.5}(d), when $\aleph =5; \nu /\Omega
=0.2$.

The curves of the SP damping in the series of  pictures in
Fig.~\ref{Fig.5} are qualitatively similar. The negative damping of the SP
becomes essential and is of order $\nu/\Omega$ near the CR, where the SP
is drastically decelerated.  The damping of the ASP is sharply diminished
in the vicinity of the point where ${\rm Re}\,p=0$, and turns to zero. The
damping of the principal SP mode  ($\omega=kv_d$)   becomes  vanishingly
small when the relaxation frequency $\nu$ is increased. This is due to the
fact that the 2DES has a small conductivity (it transforms to a
dielectric) and the surface wave is separated from the 2DES to  become a
quasibulk mode.
\section{{\bf Conclusion}} \label{four}
 To conclude it is well to  emphasize that the phase
 velocity of the SP is a remarkably small value near the CR. In other
 words, the 2D electronic layer under the QHE condition is an effectively
 decelerating system. This fact can be used for  various applications in
   microelectronics. For example, it  can be used for the
  excitation of surface electromagnetic waves by a beam of charged
 particles passing near a 2D electronic layer and for  efficient
 conversion of the beam energy into the energy of waves.

\vspace{1cm}
\centerline{{\bf ACKNOWLEDGMENTS}}
\vspace{0.5cm}

This work was supported in part by the INTAS grant No: 94-3862,
and by grants No. U2K200 and U35200 from the International Science
Foundation, and by Ukrainian Committee of Science and Technology
(project No. 2.3/19 "Metal").

\begin{figure}
\caption{The geometry of structure of 2D electronic layer
embedded in dielectric medium with dielectric constant $\varepsilon$.
\label{Fig.1}}
\end{figure}

\begin{figure}
\caption{The Hall ($\rho_{xy}$) and
longitudinal ($\rho_{xx}$) resistance as a function of magnetic field
$\bf B$  for the
typical parameters of 2D electronic structure GaAs/GaAlAs.
 \label{Fig.2}}
\end{figure}

\begin{figure}
\caption{
The lineshape of cyclotron resonance in 2DES under the QHE condition.
\label{Fig.3}}
\end{figure}

\begin{figure}
\caption{The dispersion curves for the surface polariton on the boundary
of 2DES calculated for the various values of the Landau-level filling
factor $\aleph$ ($\aleph=1; \aleph=5$, and $\aleph=10$).
 \label{Fig.4}}
\end{figure}

\begin{figure}
\caption{SP and ASP spectrum ($\omega^{'}$) (solid
line); SP damping ($\omega^{''}$) (dashed line); and
ASP damping ($\omega^{''}$)  for  various values of $\aleph$
and  $\nu/\Omega$:
\label{Fig.5}}
\end{figure}


\begin{references}
\bibitem {r15} In this model we are neglecting a number of very
interesting phenomena, e.g., Shubnikov-de Haas oscillations and the proper
description of the lineshape,  as a function of the magnetic field for the
 longitudinal d.c. conductivity $\sigma_{xx}$ and resistance $\rho_{xx}$
 under the QHE condition. The energy dependence of the relaxation
 frequency $\nu$ also can  change the CR lineshape. But in this paper we
 are interested in the high frequency ($\omega \gg \nu$) collective wave
 which shows up mainly near the CR.
\end{references}
\end{document}